\documentclass{aa}  

\usepackage{graphicx}

\usepackage{txfonts}
\usepackage{tikz,hyperref}
\definecolor{lime}{HTML}{A6CE39}
\DeclareRobustCommand{\orcidicon}{%
        \begin{tikzpicture}
        \draw[lime, fill=lime] (0,0) 
        circle [radius=0.16] 
        node[white] {{\fontfamily{qag}\selectfont \tiny ID}};
        \draw[white, fill=white] (-0.0625,0.095) 
        circle [radius=0.007];
        \end{tikzpicture}
        \hspace{-2mm}
}

\foreach \x in {A, ..., Z}{%
        \expandafter\xdef\csname orcid\x\endcsname{\noexpand\href{https://orcid.org/\csname orcidauthor\x\endcsname}{\noexpand\orcidicon}}
}

\begin{document} 

 \titlerunning{The impact of large-scale structure on the anisotropic quenching of satellites}

   \title{The impact of large-scale structure on the anisotropic quenching of satellites}


   \author{D. Zakharova\inst{1,2}\orcidA{},
          S. McGee\inst{3}\orcidD{}, 
        B. Vulcani\inst{2}\orcidB{},
         G. De Lucia\inst{4,5}\orcidC{}
             }
    \institute{Dipartimento di Fisica e Astronomia Galileo Galilei, Universit\`a degli studi di Padova, Vicolo dell’Osservatorio, 3, I-35122 Padova, Italy
    \and
    INAF – Osservatorio astronomico di Padova, Vicolo dell’Osservatorio, 5, I-35122 Padova, Italy
    \and
    School of Physics and Astronomy, University of Birmingham, Birmingham, B15 2TT, UK 
    \and
    INAF – Osservatorio Astronomico di Trieste, Via Tiepolo 11, I-34131 Trieste, Italy
    \and
    IFPU - Institute for Fundamental Physics of the Universe, via Beirut 2, 34151, Trieste, Italy}
   \date{Received September 18, 2024; accepted December 6, 2024}

 \authorrunning{Zakharova et al. }
 
  \abstract
 {Galaxies within groups exhibit characteristics different from those of galaxies that reside in regions of average density (the field). Galaxy properties also depend on their location within the host structure and orientation with respect to the central galaxy: galaxies in the inner regions that are aligned to the major axis of the central galaxy tend to be more quenched and redder than galaxies in the outskirts and with random orientation.  This phenomenon, called anisotropic satellite galaxy quenching (ASGQ), can be explained in two different ways: invoking either external influences (large-scale distribution of matter) or internal factors (black hole activity of the central galaxy). In this work, we study the impact of filaments in shaping the ASGQ in the local Universe, exploiting the magneto-hydrodynamic (MHD) simulation IllustrisTNG. We separated all surviving satellites into young and old populations depending on their infall times. We show that only young satellites contribute to the observed ASGQ. These satellites preferentially infall along the major axis of the central galaxy, which tends to have the same direction of the filament feeding the groups. We demonstrate that old satellites were quenched inside their hosts and do not exhibit signatures of ASGQ. We show that the ASGQ emerges at the time of the infall of the young satellites and is also visible outside $\rm R_{200}$. In contrast, there is no sign of anisotropic distribution in the inner regions ($R<0.5 R_{200}$). We argue that our results support a scenario in which a large-scale structure is imprinted on the ASGQ. }
 


   \maketitle
%
\section{Introduction}
\label{section:intro}

For decades, there has been an ongoing discussion regarding the distinct properties exhibited by galaxies found in high-density regions such as groups and clusters as opposed to those in the field. Pioneering work \citep[e.g.][]{Oemler+1974, Dressler+1980_morph}  showed that galaxies in clusters tend to have earlier morphological types, while disc galaxies are preferentially found in the field. Many subsequent studies revealed the importance of the environment in the evolution of galaxies~\citep{Kauffmann+2004, Cooper+2012, Vulcani+2011, Vulcani+2012, Fossati+2017}. For instance, galaxies within groups and clusters were shown to have reduced star formation rates~\citep{Peng+2010, Woo+2013, Vulcani+2010, Finn+2023} and smaller cold gas reservoirs~\citep{Giovanelli+1985, Brown+2017, Zakharova+2024} than field galaxies of similar mass. 

From a theoretical point of view,  a natural distinction can be made between central galaxies, typically located at the centre of halos (and the remaining galaxies) called satellite galaxies. These galaxies have fundamentally different evolutionary paths, as their evolution is affected by different mechanisms than central galaxies. Accretion of other galaxies, gas cooling, and active galactic nuclear~(AGN) feedback are expected to be the main mechanisms driving the evolution of central galaxies~\citep{Croton+2006, De_Lucia+2007}. The same mechanisms also contribute to the evolution of satellites, but other external mechanisms, which more strongly depend on the `environment' in which galaxies live, are also at play ~\citep{Boselli+1995, Donnari+2021, Xie+2020}. These include interactions of the galaxies with the cluster potential, other cluster members, and the intra-cluster medium, which mainly results in suppressing or halting the inflow of cold gas~(starvation or strangulation, ~\citealt{Larson+1980}), as well as gravitational disturbances~(tidal interaction,~\citealt{Bekki+1998}, or harassment,~\citealt{Farouki+1981}) or the removal of cold gas from the galaxy by stripping due to ram pressure~\citep{Gunn+1972}. These processes entail differences in the properties of central and satellite galaxies. For instance, \cite{Catinella+2013} showed that satellite galaxies have significantly lower gas content than centrals with the same stellar mass. Besides, the gas-phase metallicity of satellites is higher than centrals with similar stellar mass, which can also be explained by environmental processes~\citep{Pasquali+2013, Schaefer+2019}.
\par
Satellite galaxies may first enter small groups and be pre-processed before falling into a cluster~\citep{McGee+2009, De_Lucia+2012, Donnari+2021}. The main way to deliver groups~(also gas, galaxies, etc.) to clusters might be through cosmic filaments, which connect massive clusters~(e.g. ~\citealt{Codis+2018_connectivity}) to each other like pearl necklaces ~\citep{Tempel+2014_perls}. Moreover, neither halos nor their galaxies are randomly oriented with respect to the filaments.   N-body and hydrodynamic simulations have shown that the spin of low-mass halos is aligned with the filament axis, while that of massive halos tends to be orthogonal to the filament axis~\citep[e.g.][]{Aragon-Calvo+2007, Dubois+2014, Codis+2018_aligment, Kraljic+2020}. The same mass-dependent orientation of the spin is found for satellite galaxies in observations~\citep{Tempel+2013_aligment, Ganeshaiah+2019, Barsanti+2022}, while central galaxies seem to follow the spin orientation of their halos~\citep{West+2017, Ragone-Figueroa+2020}. 

Moreover, satellite galaxies are found to be anisotropically distributed with respect to the orientation of their central galaxies~\citep{Brainerd+2005, Yang+2006},  which tends to be aligned with that of their parent halos and the filament axis~\citep{Libeskind+2005, Zentner+2005}. Several studies \citep{Yang+2006, Martin-Navarro+2021, Stott+2022, 
Ando+2023} have demonstrated an increased amount of red and quiescent galaxies or statistically more quiescent satellites along the major axis of their central galaxy. \cite{Martin-Navarro+2021} dubbed this phenomenon anisotropic satellite galaxy quenching~(ASGQ).
\cite{Yang+2006} and \cite{Kang+2007} used a semi-analytical model to explain the ASGQ by the fact that quenched galaxies infall on the host halo along preferential directions~(e.g. along filament structures). \cite{Martin-Navarro+2021} instead explored an alternative scenario where the trend is caused by the impact of the activity of the massive black hole hosted by the central galaxy. Exploiting the IllustrisTNG hydrodynamical simulations~\citep{Illustris1, Illustris2, Illustris3, Illustris4, Illustris5}, they compared two models with similar initial conditions for the formation of large-scale structures but different treatments of AGN feedback~(an updated model that achieves realistic sufficiently rapid reddening of moderately massive galaxies and a two-mode less efficient AGN feedback scheme, see
\citealt{Weinberger+2017, Pillepich+2018} for details). Only the model with improved AGN feedback could reproduce the strength of the ASQG observed in the SDSS data \citep{SDSS}. Therefore, \cite{Martin-Navarro+2021} concluded that the AGN feedback is the main responsible for the modulation of quiescent satellites. More recently, \cite{Karp+2023} demonstrated that the  ASGQ could be reproduced in the model~\citep[UniverseMachine galaxy formation model,][]{Behroozi+2019}  without AGN feedback~(without assuming any interaction between SMBH outflows
and the CGM), and concluded that the ASGQ could instead be explained by the anisotropic accretion of galaxies onto halos.

The main purpose of this paper is to investigate the
contribution of the large-scale structure to ASGQ in groups at $z\sim0$. 
We are interested in 
revisit the analysis of the IllustrisTNG carried out by \cite{Martin-Navarro+2021} in an attempt to quantify the relative role of large-scale structure and AGN feedback.
We aim at reproducing the \cite{Martin-Navarro+2021} results, so we mimic their sample selection and fitting procedures. We complement their finding with an analysis of the orientation of satellites infall~(accretion onto the groups) using the same TNG100-1 model. 

The outline of the paper is as follows. In \hyperref[section:data]{Sect. 2,} we describe the data used. In \hyperref[section:results]{Sect. 3,} we 
present an analysis of the role of filaments in contributing to ASGQ. We inspect the distribution of infall times onto groups distinguishing between two populations of satellites. Next,  we characterise the orientation of filaments with respect to the major axis of the central galaxies. In \hyperref[sec:discussion]{Sect. 4,} we discuss the role of large-scale structure in the anisotropic satellite galaxy quenching phenomena. In \hyperref[sec:discussion]{Sect. 5,} we present our conclusions.

\section{Data and methods}
\label{section:data}
In this paper, we use the publicly available catalogues based on the TNG100-1 simulation~\citep{Illustris1, Illustris2, Illustris3, Illustris4, Illustris5} to retrieve the information presented and discussed below. TNG100-1 is a magneto-hydrodynamical (MHD) simulation corresponding to a comoving volume of $75^{3}~h^{-3}$ Mpc$^{3}$ and a baryonic particle mass of $m_{b} \sim 1.4 \cdot 10^{6}~\rm{M}_{\sun}$. It adopts a $\Lambda$CDM cosmology  with parameters: $\Omega_{\Lambda} = 0.6911$,  $\Omega_{m} = 0.3089$, $\Omega_{b} = 0.0486$, $H_{0} = 67.74$ km sec$^{-1}$ Mpc$^{-1}$, $\sigma_{8} = 0.8159$, and $n_s$ = 0.9667 \citep{Planck_Collaboration+2016}. 
\par
Following the work by \cite{Martin-Navarro+2021}, we considered all simulated halos with $12 < \log_{10}[\rm{M}_{\rm{200}, \rm{crit}} / \rm{M}_{\sun}]  < 14.2$~(hereafter $\rm{M}_{\rm{halo}}$) at $z \sim 0$. We considered only those halos that include at least two gravitationally bound subhalos~(galaxies), one of which is central~(the relation between halo and subhalo is provided in the database). All galaxies with $\log_{10}[\rm{M}_{\star} / \rm{M}_{\sun}] > 8$ inside a sphere with radius $R_{200, \rm{crit}}$~(hereafter, $\rm R_{200}$) were included in the analysis. We note that with this approach, we excluded gravitationally bound satellites beyond  $\rm R_{200}$. 
In total, the TNG100-1 simulation contains 1657 halos with central galaxies of  $ 9.7 \lesssim \log_{10}[\rm{M}_{\star} / \rm{M}_{\sun}] \lesssim  12.2$  mass; henceforth, we refer to these  halos as groups.

For each galaxy~(or subhalo in terms of the model), we extracted its position  (\texttt{SubhaloPos}), the total mass of stars bound to it ~($\rm{M}_{\star}$, \texttt{SubhaloMassType})
 and the instantaneous star formation rates~(SFR) of all gas cells in this subhalo (\texttt{SubhaloSFR})
. We identified the quiescent galaxies as  galaxies lying more than 1 dex below the star formation stellar mass sequence~(SFMS). The SFMS was estimated using all galaxies at $z \sim 0$ with  $\log_{10}[\rm{M}_{\star} / \rm{M}_{\sun}] > 8$ and $SFR > 0$ in the simulated volume\footnote{Our SFMS is largely in agreement wiht that in \cite{Martin-Navarro+2021}.}. The quiescent fraction F$_q$ is obtained by simply dividing the number of quiescent galaxies by the total number of galaxies above the same mass limit. In addition, we estimated the 3D local density of the satellites above the same mass limit. We used the volume that contains the five closest neighbors $ \displaystyle n_5 = \frac{4}{3} \pi d_{5}^3 $, where $d_{5}$ is the distance to the fifth closest galaxy with $\log_{10}[\rm{M}_{\star} / \rm{M}_{\sun}] > 8$.

We note that following \cite{Martin-Navarro+2021}, we included all satellites down to $\rm{M}_{\rm \star} / \rm{M}_{\odot} > 8 $ even though these low-mass galaxies might be artificially quenched in the simulations. This could have implications on how strong this effect might indeed be in the real Universe. 
\par
Finally, we determined the orientation of the photometric major axis of the central galaxies,  using the TNG100 catalogue of SDSS-like synthetic images with 2D Sersìc fits \citep{SKIRT_for_TNG100}.  For each galaxy with $\log_{10}[\rm{M}_{\star} / \rm{M}_{\sun}] > 9.5$~(i.e. above the mass limit we adopted for centrals) Sersìc fitting profile has been measured on synthetic \textit{g} and \textit{i} band images at $z\sim0$. 
We measured the orientation of the major axis of the galaxy from the profiles in  \textit{g} band.
If the central galaxy did not have a reliable Sersìc fitting~(S/N > 2.5 and flag$\_$sersic == 0),  we discarded the corresponding halo.
After this step, we were left with 1555 halos hosting 8162 satellites.  This number fell to 1406 after discarding all halos without satellite galaxies.


\section{Results}
\label{section:results}
\subsection{Infall times of satellites}
\label{section:infall_time_defs}

\begin{figure*}
    \centering
    \includegraphics[width=1\linewidth]{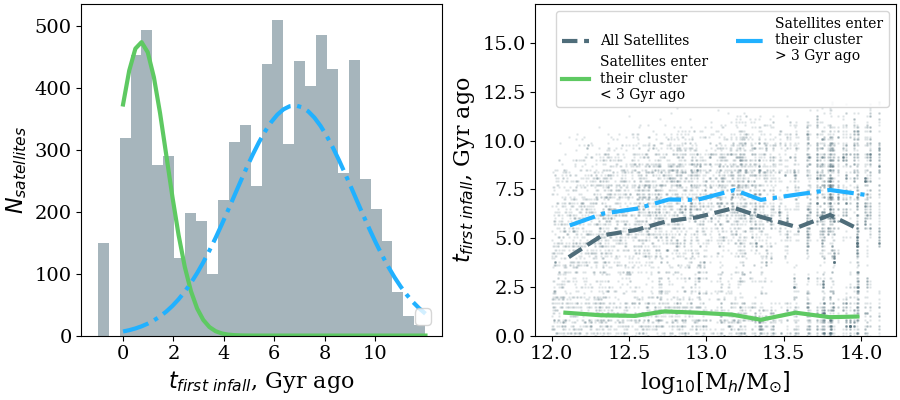}
    \caption{First infall time of satellite progenitors onto their host-group progenitor (see text for details). \textit{Left:}  Probability distribution function of  $t_{first~infall}$. The Gaussians illustrate the presence of two populations of satellites: those whose progenitors first entered before 3 Gyr ago~(old population, in blue) and those whose progenitors first entered in the last 3 Gyr~(young population, in green). Negative values represent satellites whose first infall time could not be computed~(empty merger tree for low-mass subhalos, near the resolution limit, for instance). \textit{Right:} First infall time as a function of host halo mass $\rm{M}_{\rm halo}$ at z$\sim$0. Lines represent the median values for the old~(blue line), total~(grey line), and young~(green line) populations of satellites.  }
    \label{fig:time_of_entrance}
\end{figure*}

In this section, we aim to establish whether satellites have a preferential orientation at the time of infall and whether there is evidence of  ASGQ when considering galaxies before they infall onto groups. We define $t_{first~infall}$ as the time corresponding to the snapshot in which the main progenitor of each surviving\footnote{We define surviving satellites as those whose descendants can be identified as satellites at $z\sim0$.} satellite first crosses the virial radius, $\rm R_{200}$, of the main progenitor of the parent halo at $z\sim0$. The main progenitors of each galaxy were identified by the \texttt{SubLink algorithm}~(more details can be found in \citealt{Sublink_alg}).

The left panel of Fig.~\ref{fig:time_of_entrance} shows the distribution of the first infall time for satellites into their groups at $z\sim0$. The distribution shows a clear bimodality 
with a  minimum of around 3 Gyr ago so that we can identify two distinct populations: 
\begin{itemize}
    \item a young population of satellites, whose progenitors firstly enter the host-group less than 3 Gyr -- 29$\pm$1\% of the total number of satellites,
    \item an old population of satellites, whose progenitors first infall onto the host group more than 3 Gyr -- 71$\pm$1\% of total number of satellites.
\end{itemize}
The physical origin of this bimodality remains unclear. However, this feature has been identified in other studies based on different approaches, such as the semi-analytic model GAEA~\citep{De_Lucia+2012} and N-body simulations in \cite{Oman+2016}. 


The right panel of Fig.~\ref{fig:time_of_entrance} shows $t_{first~infall}$ as a function of halo mass. The figure shows median values in bins of $\rm{M}_{\rm halo}$  for all satellites and the old and young populations separately. The median value for the old population is about 6 Gyr when considering low-mass halos ($\rm{M}_{\rm halo} \sim 10^{12} \rm{M}_{\sun}$), and rises to about 8 Gyr for massive halos ($\rm{M}_{\rm halo} \sim 10^{14} \rm{M}_{\sun}$). In contrast, the young population does not show a dependence of $t_{first~infall}$ on the mass of the halo.

\begin{figure*}
    \centering
    \includegraphics[width=1\linewidth]{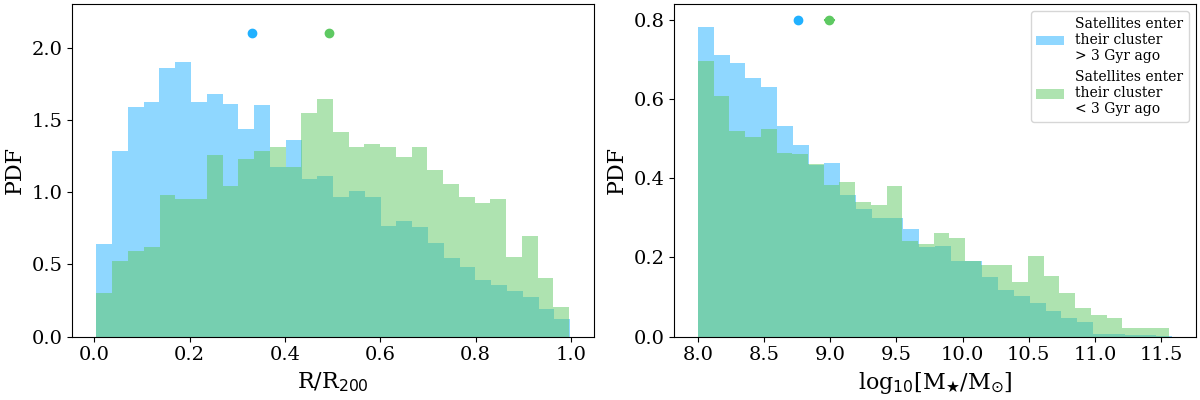}
    \caption{Spatial and mass properties of old~(blue) and young~(green) satellite populations. \textit{Left:} Normalised radial distribution, in units of R$_{200}$. \textit{Right:} Stellar mass distribution of galaxies in the young and old satellite populations. The median values and 1$\sigma$ confidence interval are shown~(small).
    }
    \label{fig:radii_andn5}
\end{figure*}

We now proceed to characterise the properties of the satellites in the two populations identified above. 
First, we are interested in the spatial distribution of the old and young satellites. For each galaxy, we calculated the 3D distance to the central galaxy and normalise it to $\rm R_{200}$ at $z\sim0$. The left panel of Fig.~\ref{fig:radii_andn5} shows that old satellites are most likely located in the inner parts of their groups~(the median radii with 1 $\sigma $ confidence interval is 0.33$\pm$0.01), while young satellites tend to be located in the outer regions~(the median radius is 0.49$\pm$0.01). This is a natural consequence of dynamical friction: Satellites that were accreted earlier have had more time to spiral towards the inner regions of the halo~\citep{De_Lucia+2012}. 

Next, we investigated whether there is a difference in the stellar mass distribution of the two populations. The right panel of Fig.~\ref{fig:radii_andn5} shows that distributions are different: there is a larger fraction of more massive galaxies in the young rather than in the old population: 9$\pm$1\% of the galaxies in the young population have $\log_{10} [\rm M_{\star} /\rm{M}_{\sun} ] > 10.5,$ while this fraction drops to 3$\pm$1\% when considering the old population. Even though the median values are not significantly different, the Kolmogorov-Smirnov test supports the finding that the distributions are drawn from different parent samples. 

This finding is a consequence of the fact that galaxies that were accreted later have more time to grow.  Also, the most massive satellites that have been accreted earlier on have likely already merged due to their short merger times. Therefore, the small difference just discussed could be the selection effect~\citep{De_Lucia+2012}. 

In summary,  surviving satellites can be divided into an old and a young population. The former is made of galaxies that entered their host group more than 3 Gyr ago and tend to be located in the inner radii of their parent halos. The latter is made of satellites accreted in the last 3 Gyr and their spatial distribution is skewed towards the outskirts of their parent halos. Using this result, we went on to inspect the ASGQ for these two separate populations and their orientation at the time of accretion.

\subsection{Anisotropic satellite quenching}
\label{sec:asq_intro_by_gens}

Following \cite{Martin-Navarro+2021}, we computed the modulation of the quiescent fraction of satellite galaxies located in groups. We estimated the orientation of each satellite galaxy relative to the orientation of the major axis of their central galaxy at $z\sim0$. We first calculated the position of each satellite relative to the central galaxy in the  XY-plane:
 \begin{equation}
        \begin{cases}
         x_{sat,0} = x_{sat} - x_{cen}, \\
          \\
            y_{sat,0} = y_{sat} - y_{cen},
        \end{cases}\,
        \label{eq:sat_orient1}
\end{equation}
and then transform the coordinates from cartesian to polar:
 \begin{equation}
        \begin{cases}
         \theta_{sat} = \arctan \frac{y_{sat,0}}{x_{sat,0}} \\
          \\
            r_{sat} = \sqrt{(x_{sat, 0}^2 + y_{sat, 0}^2 )}
            \label{eq:sat_orient2}
        \end{cases}\,
    ,\end{equation}
where $\theta$ is the angle between the direction of the satellite and the positive direction of the X-axis. Then, the orientation of the satellites relative to the major axis of the central galaxy is $\theta = \theta_{sat}$ - $\theta_{cen}$. Thus, satellites along the major axis of their central galaxy have an orientation of $\theta \sim 0$ deg or $\theta \sim 180$ deg, while along the minor axis, the orientation is $\theta \sim 90$ deg or $\theta \sim 270$ deg.

\begin{figure*}
    \centering
    \includegraphics[width=1\linewidth]{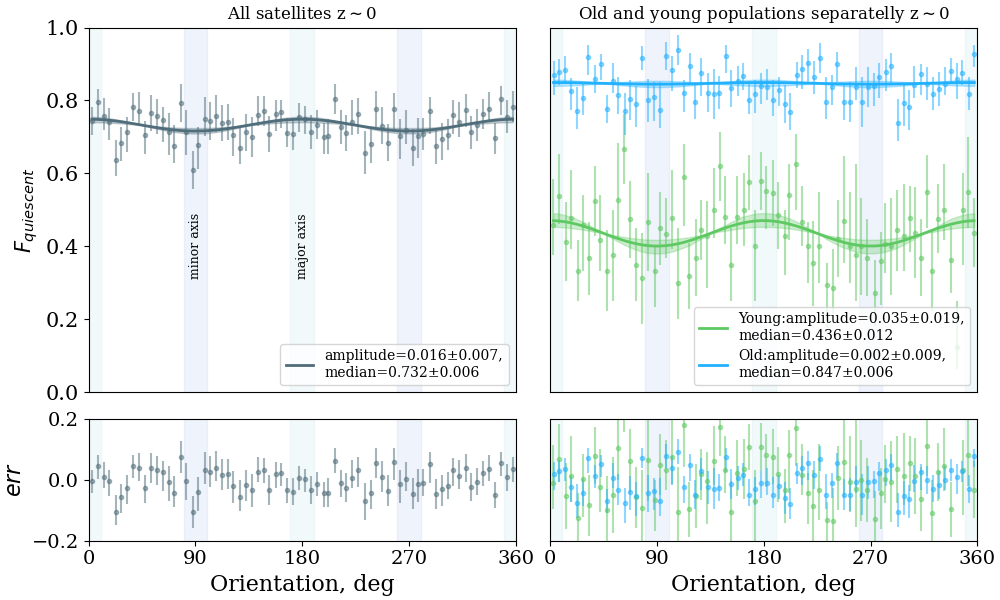}
    \caption{Modulation of the quiescent fraction of satellite galaxies as a function of orientation with respect to the major axis of their central galaxy. Symbols with error bars show the median values with 1$\sigma$ confidence interval in bins of 5 deg. The modulation is fitted by a cosine function, and the amplitude of the modulation and median $\rm F_{\rm{q}}$ are indicated in the right bottom corner with a 1$\sigma$ confidence interval. The shaded zone around the cosine function indicates a 1$\sigma$ confidence interval in the amplitude fitting.  The bottom panels represent the residuals between the data and the fitting. The orientation of the minor and major axes of central galaxies is indicated by vertically shaded regions. \textit{Left:} Results for all 8162 satellites $\log_{10}[\rm{M}_{\star}/ \rm{M}_{\odot}] > 8$ inside $\rm R_{200}$  at z$\sim$0. The marginal distributions of the fitting parameters are shown in Fig.~\ref{fig2_fitting1}. \textit{Right:} Same, but for young~(green line) and old ~(blue line)  satellites separately. The corresponding marginal distributions of the fitting parameters are shown in Fig.~\ref{fig:fig2_fitting_young} and \ref{fig:fig2_fitting_old}.}
    \label{fig:asgq_z0}
\end{figure*}

Using the orientations of all satellites with respect to the central galaxy of their halos, we grouped galaxies in bins of 5 degrees. For each of these bins, we obtained the median values and the  $1\sigma$ confidence interval estimated by bootstrapping the quiescent fraction $\rm{F}_{\rm{q}}$ and the local density $\rm n_5$, both defined in Sect. 2. We introduced the local density as a proxy for large-scale structure since our main goal is to test the contribution of the large scale structure to the modulation of $\rm{F}_{\rm{q}}$. The results obtained for all halos stacked together are shown in the left panel of Fig.~\ref{fig:asgq_z0}.

We fit the quiescent fraction $\rm{F}_{\rm{q}}$  as a function of the orientation with a cosine function using a Markov chain Monte Carlo (MCMC) sampler to evaluate the following likelihood function, as given in \cite{Martin-Navarro+2021}:
\begin{align}
    \ln p(f_{q}| \theta, a, b, f) = & - \frac{1}{2} \sum_{i} \frac{(f_{q,i} - a - b \cos 2\theta_i)^2}{\sigma^2 + f^2} + \notag \\
    & + \ln [2\pi (\sigma^2 + f^2)],
    \label{eq:fitting}
\end{align}
where $f_{q,i}$ is the observed fraction of quiescent galaxies at the $\theta_i$ orientation, \textbf{a} is the median quiescent fraction, and \textbf{b} is the amplitude of the modulation. The uncertainty is given by $s^2 = \sigma^2 + f^2$, where $\sigma$ is the estimated error, in our case estimated by bootstrapping, and $f$ is the rescaling term.
The result of the fitting procedure is also shown in Fig.~\ref{fig:asgq_z0}~(left panel). Overall, we recovered the modulation reported in \cite{Martin-Navarro+2021}: satellites aligned with the major axis of the central galaxy have a higher probability of being quiescent than satellites along the minor axis. The median quenched fraction is  $\rm F_{\rm q} = 0.73\pm0.01$ and the amplitude of the modulation is of $0.016$ dex, which is slightly lower than that in \cite{Martin-Navarro+2021}.

\cite{Martin-Navarro+2021} were not clear about the satellites used in their work. Most probably, they considered all gravitationally bound satellites, whereas we applied a cut at $\rm{R}_{200}$. If we relax this constraint,  we obtain $\rm F_{\rm q} = 0.64\pm0.01$ and an amplitude of $0.029$, which is in agreement with \cite{Martin-Navarro+2021} result.


Next, we investigated whether old and young satellites separately show the same  ASQG signal. 
As before, we estimate the quiescent fraction in the bins of  5 degrees and estimate uncertainties by bootstrapping. The right panel of Fig.~\ref{fig:asgq_z0} shows  that satellites that entered their parent halo more than 3 Gyr ago have significantly higher quiescent fractions than young satellites: 
$\rm F_{\rm{q,~old}} = 0.85\pm0.01$ and $\rm F_{\rm{q,~young}} = 0.44\pm0.01$. Thus, satellite galaxies that spend more time inside their group reside in the inner regions, and a larger fraction of them are quenched, as expected (e.g. ~\citealt{DeLucia+2004,  De_Lucia+2012, George+2013, Vulcani+2015}).

Figure~\ref{fig:asgq_z0} also shows that old satellites within $\rm R_{200}$ do not show any modulation of the quiescent fraction as a function of orientation~(the amplitude is close to 0). We argue that any modulation imprinted at the time of satellite accretion is lost during their subsequent evolution within the groups. Besides, the central galaxy changes its orientation significantly and even abruptly, which could also blur any existing modulation (see Appendix~\ref{app:centrals_orientation_over_time}). Unlike the old population, young satellites inside $\rm R_{200}$  show a pronounced modulation, with an amplitude of $0.035$ dex. These results show that the ASGQ signal is driven by the young satellites.

In summary, old and young satellite populations exhibit different properties of ASGQ: only recently accreted satellites show a strong modulation. In contrast, the old populations are characterised by a much higher median $F_q$, comparable to the median $F_q$ of satellites at $ z\sim0$, but they exhibit no modulation.

We note that through this analysis, we used instantaneous star formation rates of galaxies to closely follow the analysis by \cite{Martin-Navarro+2021}. However, time-averaged star formation rates better align with values typically derived from observational measurements.  To assess the impact of this definition on our results, we performed our analysis again using star formation rates obtained by averaging the rates on the last 100 Myr and considering all gravitationally bound stars within the subhalo. These data are available in the catalogues of \cite{Donnari+2019} and \cite{Pillepich+2019}. Qualitatively, our results remain the same: the ASGQ signal at z=0 is recovered only for the young satellites, also before the infall into their host structure. The analogues of Figs.~\ref{fig:asgq_z0} and ~\ref{fig:asgq_yong_modulation_before_entrance} with time-averaged star formation rates are shown in Appendix~\ref{app:sfr_def}.
\subsection{Orientation and  modulation of  $\rm F_{\rm q}$ at infall time for the young populations}
\label{subsec:young_gens_props}
\begin{figure*}
    \centering
    \includegraphics[width=1\linewidth]{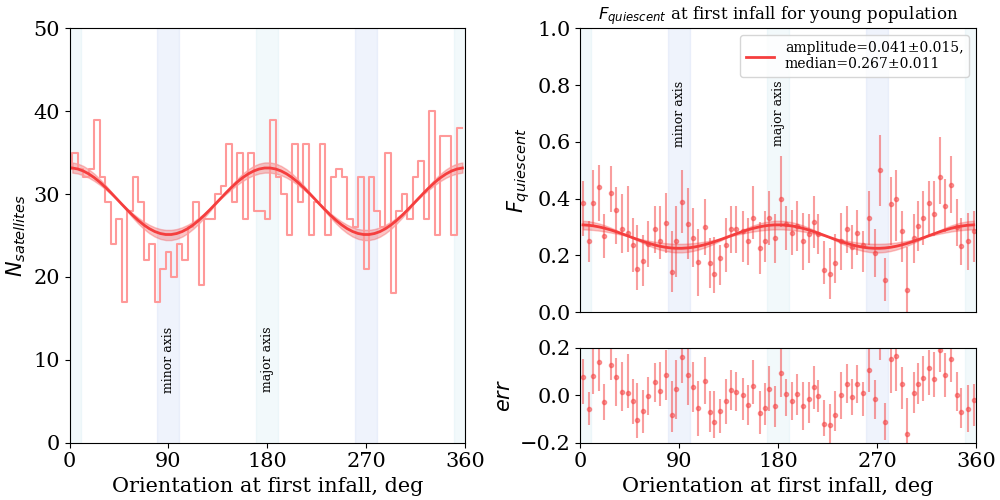}
    \caption{Modulation of infall orientations and quiescent fraction of young satellites at the time of infall. \textit{Left:} Distribution of the orientation of the progenitors of satellites in the young population at the first infall with respect to the major axis of the central galaxy~(at $z\sim0$). The solid line shows a fit to  Eq.~\ref{eq:fitting}. \textit{Right:} Modulation of the fraction of quiescent satellites progenitors of the young population at first infall time~(similar to Fig.~\ref{fig:asgq_z0}). The marginal distributions of the fitting parameters are shown in Fig.~\ref{fig:fig3_fitting}.}   
    \label{fig:asgq_yong_modulation_before_entrance}
\end{figure*}
We now investigate the properties of the satellite's progenitors at the time of infall to check whether a significant ASGQ signal can be measured before satellites interact with their host halos. In Appendix~\ref{app:centrals_orientation_over_time} we discuss how the orientation of the central galaxy changes over time. The main conclusion is that in the last 3 Gyr, the dynamical orientation of the central galaxy has barely changed, so it allows us to assume that  its photometrical orientation did not change either. Therefore, we can estimate the orientation of the young populations at the time of infall with respect to the $z\sim0$ orientation of the central galaxy. We note that this assumption is not valid for the old population, as at epochs older than 3 Gyr ago, often central galaxies changed their orientation  (see Appendix~\ref{app:centrals_orientation_over_time}).

For the young satellites, we define the orientation of the first infall by Eqs.~\ref{eq:sat_orient1} and~\ref{eq:sat_orient2} and show its distribution in the left panel of Fig.~\ref{fig:asgq_yong_modulation_before_entrance}.
The orientation of young satellites at infall also exhibits a modulation, with maxima corresponding to the direction of the major axis of the central galaxy. So, young satellites infall onto their host halos unevenly with respect to the major axis of the central galaxy. An increased number of satellites have been accreted along this axis.

Next, we wish to understand whether the ASGQ signal was already present at the first infall time for the young population of satellites. Hence, we determined if the progenitors of the satellites were star-forming or quiescent at the time of infall. As done before, at each snapshot, we compute the SFMS using all galaxies with $\log_{10}[\rm{M}_{\star}/ \rm{M}_{\odot}] > 8$ in the simulated volume and define as quiescent galaxies those whose star formation is below 1 dex from the corresponding relation. First of all, we measure the median quiescent fraction of the young satellites at first infall, which is $\rm{F}_{\rm q}=0.27\pm0.01$. If we measure the same quantity for the old population, we find that only 9\% of their progenitors were quenched at the time of first infall, suggesting that old satellites preferentially fell onto their groups as star-forming galaxies\footnote{Since around 70\% of surviving satellites are quenched, this is in perfect agreement with \cite{Martin-Navarro+2021}, who claimed that the vast majority of satellites were quenched within their groups. }. 

\indent We then obtain the modulation of $\rm{F}_{\rm q}$ as a function of the orientation at infall time of young satellites fit the data using the  Eq.~\ref{eq:fitting}. 
The amplitude of the modulation we measure at first infall time is 0.041$\pm$0.015 dex, which is in agreement with $z\sim0$~(0.035$\pm$0.02). We interpret the existence of the $\rm F_{\rm q}$ modulation already at first infall as evidence that cosmic filaments are the main contributor to the ASGQ as was suggested in \citealt{Kang+2007} and \cite{Karp+2023}. These studies assumed that galaxies enter the groups through filaments. Therefore, in the next section we check if the filaments in TNG100-1 are aligned with the orientation of infalling young satellites, namely, along the central galaxy's major axis.
{We note that our results are based on a halo sample covering a wide mass range $12 < \log_{10} [\rm{M}_{\rm halo} / \rm{M}_{\odot}] < 14.2 $. In Appendix~\ref{app:clusters_and_grousp_separately} we show the impact of this choice on our results. }
 
\subsection{Filaments orientation relative to the major axis of the central galaxy}
\label{sec:filaments_orient}

\begin{figure}
    \centering
    \includegraphics[width=1\linewidth]{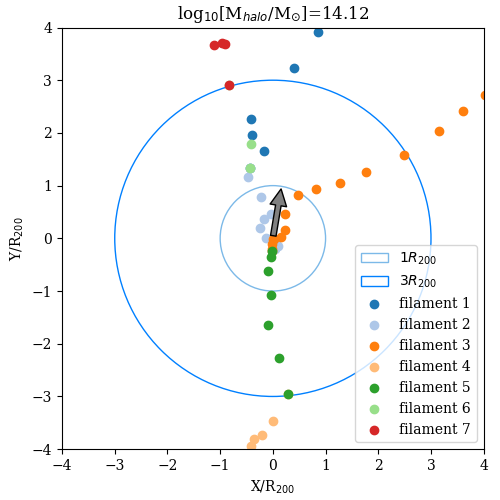}
    \caption{Example of the arrangement of filaments within 1 and 3 virial radii $\rm R_{200}$ of the halo around a group with $\log \rm M_{\rm halo} = 14.2$.  
    The position of filament points in a cubic box of $(8\rm R_{200})^3$ is shown by coloured points. The circles indicated 1 and 3$\rm R_{200}$, respectively. The grey arrow shows the orientation of the major axis of the central galaxy. Within 1$\rm R_{200}$, filaments 2, 3, and 5 are aligned with the central galaxy orientation.  }
    \label{fig:filaments_orient_example}
\end{figure}
Here, we verify 
whether the major axis of central galaxies in the TNG100-1 simulation is oriented along the filaments as shown in other works~(e.g. ~\citealt{Tempel+2013_aligment, Zhang+2013} using SDSS or \citealt{Codis+2018_aligment} based on the hydrodynamical simulation Horizon-AGN). We first extracted filaments considering all galaxies $\log_{10}[\rm{M}_{\star} / \rm{M}_{\rm {halo} } ] > 8$ in the entire TNG100-1 simulated volume at z$\sim$0. We identify filaments using DisPerSE~\citep{Sousbie+2011, Sousbie_etal+2011} with $5 \sigma$ persistence level~(this is the typical value used in most studies using this algorithm).  In addition, we reject filaments shorter than 3 Mpc/h. We mainly followed the approach described in \cite{Zakharova+2023}, so a more detailed description of the used DisPerSE settings can be found there. DisPerSE provides filaments' spine positions as a set of points. Fig.~\ref{fig:filaments_orient_example} shows an example of one group with identified filament points within a cubic box of $(8\rm R_{200})^3$. The image also shows the orientation of the central galaxy, which in this example is aligned with all the filaments inside the halos' virial radii (i.e. filaments 2, 3, and 5).

For each considered group, we identify filaments that cross the spheres 1$\rm R_{200}$ and 3$\rm R_{200}$ and we calculated the orientation of the position of each filament point with respect to the central galaxy. We use two different radii to check whether trends change beyond the halo virial radius, as it occurred (e.g. to filament 3 in Fig.~\ref{fig:filaments_orient_example}). The following steps were taken in the procedure. 
\begin{enumerate}

    \item We determined the position of filaments points relative to the central galaxy (XY projection) similarly to the Eq.~\ref{eq:sat_orient1},
    \item We converted the positions into polar coordinates~(orientation  $\theta_{fil}$ and radii  $r_{fil}$) similarly to Eq.~\ref{eq:sat_orient2}.
    \item We measured the orientation of the
filament relative to the major axis of the central galaxy as $\theta = \theta_{fil}$ - $\theta_{cen}$.
\end{enumerate}

\begin{figure*}
    \centering
    \includegraphics[width=1\linewidth]{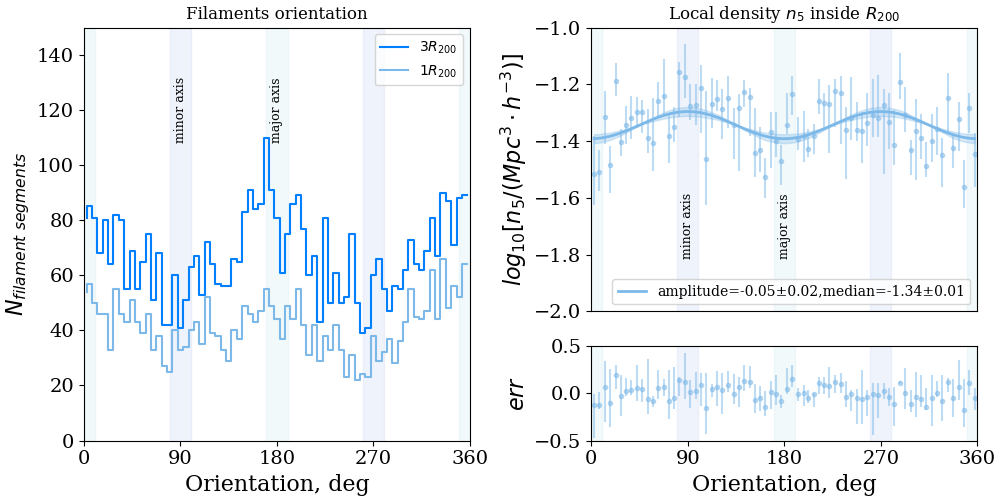}
    \caption{Modulation of orientation of filaments around 1$\rm R_{200}$ and $3R_{200}$ of halos and corresponding local density.
    \textit{Left:} The orientation of filaments points inside $\rm R_{200}$ and $3R_{200}$ with respect to the major axis of centrals for all groups. \textit{Right:} Local density $n_{\rm 5}$  inside $\rm R_{200}$ of the groups according to the major axis of the central galaxy.  }
    \label{fig:filaments_along_central}
\end{figure*}

The left panel of Fig.~\ref{fig:filaments_along_central} shows the distribution of the orientations of filament points inside 1$\rm R_{200}$ and 3$\rm R_{200}$, respectively. For both apertures considered, 
filament points tend to be aligned with the major axis of the central galaxies. 

Since the filament detection is essentially based on the identification of local overdensities~\citep{Sousbie+2011, Sousbie_etal+2011}, we can expect a modulation with respect to the orientation of the central galaxy also when inspecting the local density. Indeed, the right panel of Fig.~\ref{fig:filaments_along_central} shows that denser areas are located along the major axis of the central galaxy.
In conclusion, filaments and regions of increased density in groups are aligned with the orientation of the major axis of the central galaxy in the TNG100-1 simulation.

\section{ASGQ as a consequence of the co-orientation of central galaxies and filaments}
\label{sec:discussion}

\begin{figure*}
    \centering
    \includegraphics[width=0.8\linewidth]{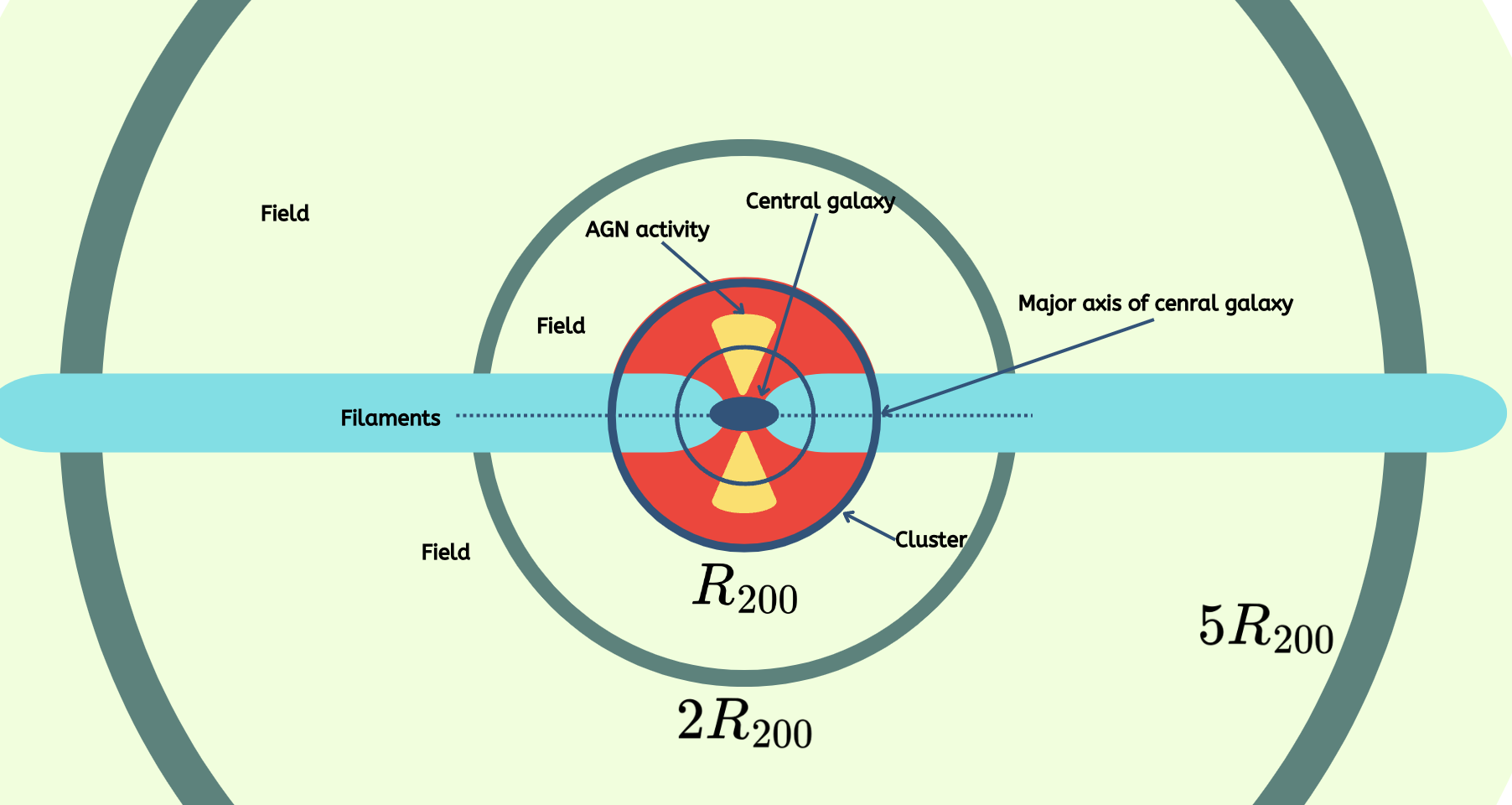}
    \caption{Schematic representation of a group connected with two filaments aligned with the central galaxy's major axis. In this case, there is a  modulation of the local density both in the inner regions of the halo~($R < R_{200}$) and outside ~($R > R_{200}$).}
    \label{fig:scame_filamets}
\end{figure*}

\begin{figure*}
    \centering
    \includegraphics[width=1\linewidth]{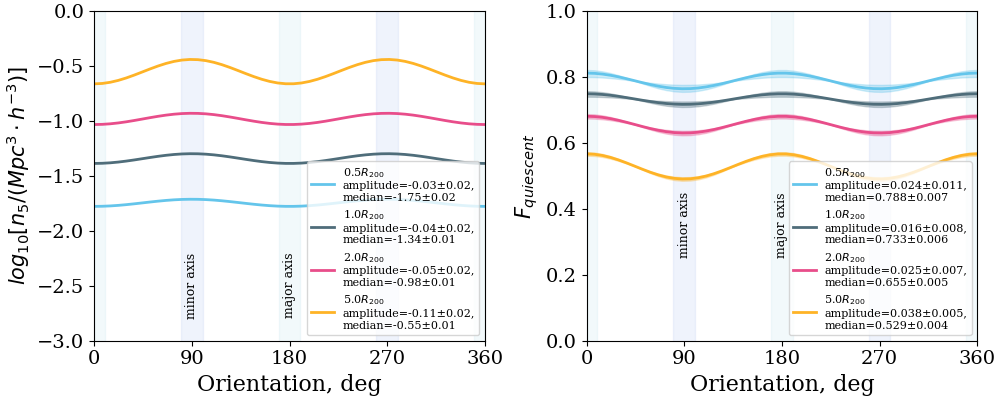}
    \caption{Modulation of local density $n_{\rm 5}$~(left panel) and quiescent fraction $\rm F_{\rm q}$~(right panel) in spheres surrounding central galaxies  with radii of 0.5$\rm R_{200}$, 1$\rm R_{200}$, 2$\rm R_{200}$, and 5$\rm R_{200}$.}
    \label{fig:asgq_diff_r}
\end{figure*}

In the previous sections, we demonstrated that the orientation of the major axis of the central galaxy
is co-oriented with the filaments and regions where satellites show higher local densities. In the same regions, we detected a higher fraction of quenched galaxies~(Fig.\ref{fig:asgq_z0}, right panel). Thus, our findings support the interpretation by \cite{Kang+2007} and \cite{Karp+2023} that the presence of the ASGQ signal due to the specific orientation of the central galaxy connected to the large-scale structure and, in particular, to the position of the filaments entering the halos.

Figure~\ref{fig:scame_filamets} schematically illustrates the possible scenario:  the major axis of the central galaxy is aligned with filaments. 
If the ASGQ phenomenon within the groups ($R < R_{200}$) is due to the influx of galaxies with a higher probability of being quenched (hence a population with a higher $\rm F_{\rm q}$) along the filaments, then, we expect the modulation of  $\rm F_{\rm q}$  even outside the groups ($R \ge R_{200}$). Such modulation should be originated by the fact that galaxies located in filaments aligned to the major axis of the central galaxy have a higher $\rm F_{\rm q}$ than field galaxies, which can fall onto halos with random orientations.

Nonetheless, in all the cases considered, the modulation of the local density is evident, with peaks along the major axis of the central galaxy. In contrast, the amplitude of the modulation increases with increasing radii. This means that the local density contrast is higher outside  R > $\rm R_{200}$ than inside. In the R > $\rm R_{200}$ case, the local density reflects the density contrast between the general field and filaments, as expected by Fig.~\ref{fig:scame_filamets}. At 0.5 $\rm R_{200}$ < R < $\rm R_{200}$, the densest regions correspond to the places where the filaments are connected to groups. The low modulation in the inner parts of the groups characterised by  R < 0.5 $\rm R_{200}$  reveals that filaments have a limited effect on the most central regions of the groups. 

Thus, the modulation of local density according to the central galaxies exists not only inside their groups but also outside and even in the groups' outskirts~(at least up to $R$ $<$ 5$\rm R_{200}$). This indicates that the orientation of the central galaxy is strongly correlated to the large-scale structure.

Similarly to the local density, we can investigate how the modulation of the quiescent fraction varies with the distance from the centre. The right panel of Fig.~\ref{fig:asgq_diff_r} shows the modulation of $\rm{F}_{\rm q}$ for galaxies within spheres 0.5, 1, 2, and 5 $\rm R_{200}$. It clearly shows that the local density is connected to $\rm{F}_{\rm q}$: the inner parts of the groups have the highest fraction of quiescent galaxies and almost no modulation, while at the larger distances from the centrals, the median $\rm{F}_{\rm q}$ decreases while the modulation increases. Inside $R < 0.5 \rm R_{200}$, around 80\% galaxies are quiescent, independent of the modulation. 
Inside  $0.5 \rm R_{200} < R < 1R_{200}$, the modulation exists and correlates with local density: the peak of $\rm{F}_{\rm q}$ corresponds to the regions, where filaments deliver a higher fraction of quenched galaxies than along the minor axis.
At larger radii, $R = 2 \rm R_{200}$ or $R = 5 \rm R_{200}$, the modulation of $\rm{F}_{\rm q}$  exists and increases with the considered volume, the difference between the quiescent fraction of the general field~(along the minor axis) and of the filaments~(along the major axis, see Fig.~\ref{fig:scame_filamets}). We argue that the modulation of the local density and of the $\rm{F}_{\rm q}$ can be regulated by the central galaxy even outside the group. 


The correlation between local density and $\rm{F}_{\rm q}$ inside $ R < 1R_{200}$ might be explained by the internal effects inside groups, such as central galaxy AGN activity not outside of the groups. In contrast, the presence of modulations~(local density and $\rm{F}_{\rm q}$) outside the groups can only be explained by the global distribution of matter within the large-scale structure, coordinated with the major axis of the central galaxy.

It is worth adding that the presence of the anisotropic satellite galaxy quenching outside the virial radius of $1R_{200}< R  <2 \rm R_{200}$ also indicates that the accretion of the next population of satellites will also be anisotropic in direction. It will take place preferentially along the filaments (along the major axis of the central galaxy), as shown for young satellites in Fig~\ref{fig:asgq_yong_modulation_before_entrance}.

\section{Conclusion}
\label{sec:conclusion}

In this paper, we analyse the properties of galaxies inside groups with $12 < \log_{10} [\rm{M}_{\rm halo} / \rm{M}_{\sun}] < 14.2 $ in the TNG100-1 simulation. We specifically inspected the connection between large-scale structures and anisotropic satellite galaxy quenching. 

We have shown that galaxies inside groups can be divided into two populations based on their infall times: old and young satellite populations. 
The vast majority of today's satellites~(70\%) were\ accreted onto the progenitors of their parent halo more than 3 Gyr ago. Today, they are located in the inner radii of the groups and are predominantly quenched~($\rm{F}_{q, old} = 0.84\pm0.01$). These 'old' satellites do not exhibit anisotropic satellite quenching. 

We note that we do not expect any signature of ASGQ for old satellites for two reasons. First,  the orbits of the satellites are complex and mix throughout the life of the group, so it is unlikely that early accreted satellites maintain a fixed and preferential orientation relative to the central galaxy. Second, as we discuss in Appendix~\ref{app:centrals_orientation_over_time}, the central galaxy also changes its orientation during its lifetime~(except the last 3 Gyr), which would blur any ASGQ signal.

For satellites accreted later than 3 Gyr ago, we obtain a pronounced modulation of the quiescent fraction $\rm{F}_{\rm q}$. We consider these 'young'  satellites to be responsible for the existence of the modulation that is evident when the entire population is considered \citep[e.g.][]{Martin-Navarro+2021}. We have shown that these satellites exhibit modulation already at infall time and have a preferential orientation of accretion along the major axis of the central galaxy~(enter their host halo anisotropically). 

We demonstrate that the ASGQ signal exists not only inside $\rm R_{200}$ but also beyond the virial radii of groups. Therefore, we argue that this signal cannot be explained only by processes at play inside the groups~(as was suggested by \citealt{Martin-Navarro+2021}) and we relate it to the non-random orientation of the central galaxies. Besides, the overdensity of quenched galaxies arises along the major axis of the central galaxy, which is co-oriented with filaments. We have shown that the direction of the filaments is also aligned with the major axis of the central galaxy, which indicates consequences of halo accretion~\citep{Karp+2023}. This result is in agreement with previous studies that explored the orientation of galaxies within the cosmic web~(\citealt{Zhang+2013, Codis+2018_aligment, Kraljic+2020} and many more).
Finally, we demonstrate that filaments impact groups inside $ 0.5 \rm R_{200} < R < R_{200}$ and have almost no influence in the innermost regions at  $ R < 0.5 \rm R_{200}$, which were detected according to the analysis of $n_{5}$.


\begin{acknowledgements}
        We are thankful to the referee whose comments and suggestions helped us to improve our manuscript.
      DZ and BV acknowledge support from the INAF Mini Grant 2022, “Tracing filaments through cosmic time”  (PI Vulcani). SM acknowledges support from UK Space Agency grants ST/Y005201/1 and ST/Y000692/1. 
\end{acknowledgements}

%
  \bibliographystyle{aa} 
  \bibliography{main} 
%
\begin{appendix}

\section{Changes in the orientation of central galaxies over time in TNG100}
\label{app:centrals_orientation_over_time}

In the main part of the work, we found that when considering the old populations of satellites,  there is no dependence of the fraction of quiescent galaxies on the orientation of the central galaxy. This could be explained, for example, by frequent and/or abrupt~(in the case of merging) changes in the orientation of the central galaxy in the past so that older populations no longer carry information about the orientation of the central galaxy. Besides, to estimate the orientation infall of young populations of satellites, we use the orientation of the central galaxy at $z\sim0$, so we need to check the foundation of this choice. 

We focus on the changes during all times of central galaxies life and in particular during the last 3 Gyr. Although the change in the orientation of central galaxies in the past is of great interest, the reasons or consequences of this result are beyond the scope of this paper. We refer, for example, to  \cite{Rodriguez+2024}. Here, we stress how the central galaxy generally changes its orientation during its lifetime.

In the main part of the work, we used the orientation of the photometrical major axis estimated by the S\'ersic index on synthetic images of the central galaxies. However, those exist only for two snapshots corresponding to the $z\sim0$ and $z\sim0.05$. Getting such data by ourselves by running the SKIRT code for each progenitor of the central galaxy would be computationally heavy. Therefore, to estimate the evolution of the orientation of central galaxies during their lifetime, we calculate the dynamical orientation of the major axis by inertia tensor. We note that the dynamic and photometric orientations of galaxies are not obviously related, and strictly speaking, they do not have to match. However, we assume that analysis of changes in a galaxy's dynamic major axis allows us to draw conclusions about the frequency of changes in the photometric major axis.

\begin{figure}
    \centering
    \includegraphics[width=1\linewidth]{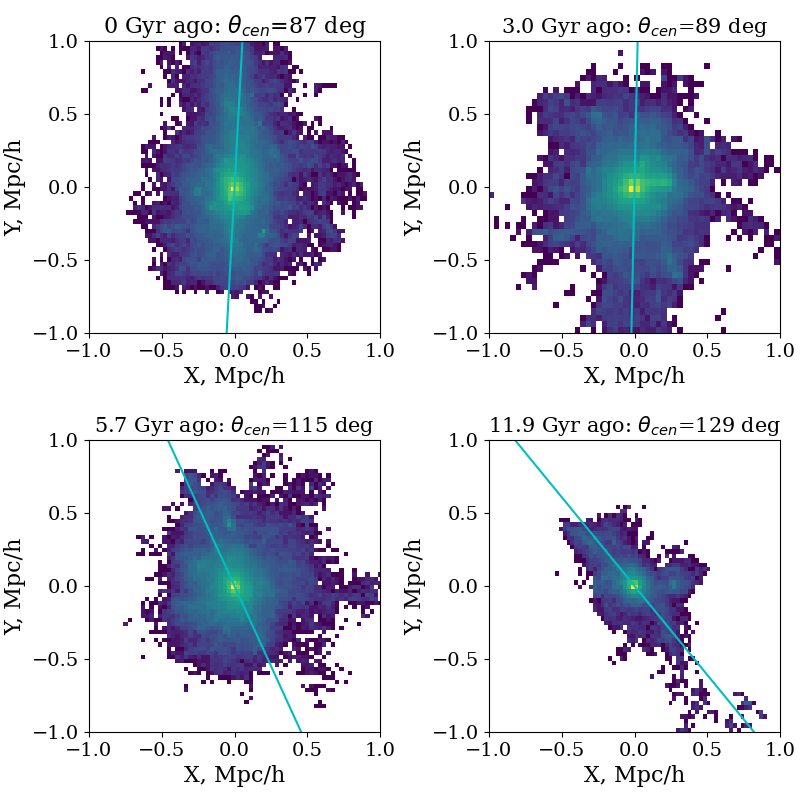}
    \caption{Evolution of orientation of central galaxy $\log_{10}[\rm{M}_{\star} / \rm{M}_{\sun} ]= 11.96$ of group $\log_{10}[\rm{M}_{\rm{halo}} / \rm{M}_{\sun } ] = 14.02$. Each panel shows a distribution of star particles of a galaxy of a given timestamp and the position of the major axis of the galaxy, respectively. }
    \label{fig:central_orient_illustration_3snaps}
\end{figure}

\begin{figure}
    \centering
    \includegraphics[width=1\linewidth]{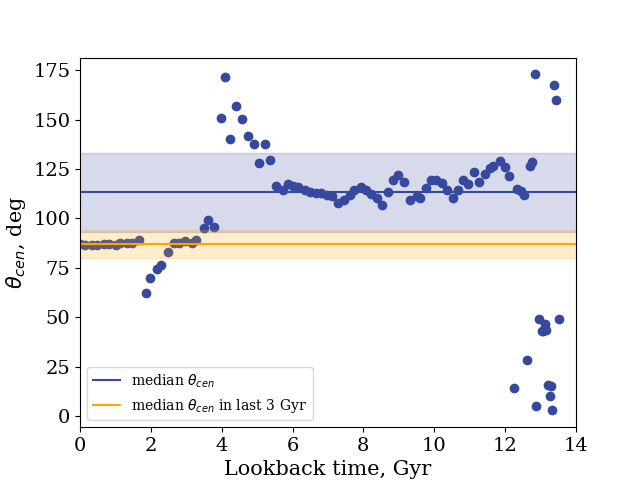}
    \caption{Orientation of the major axis of the galaxy from Fig.~\ref{fig:central_orient_illustration_3snaps} for each timestamp. Each dot represents the orientation of the major axis for every timestamp. The line and shaded zone represent the median orientation and standard deviation for all times and the last 3 Gyr.}
    \label{fig:central_orient_illustration2}
\end{figure}

The inertia tensor \( \rm I \) for star particles for each central galaxy:
\[
\rm I = \begin{bmatrix}
\sum_{i} m_i (y_i^2 + z_i^2) & -\sum_{i} m_i x_i y_i & -\sum_{i} m_i x_i z_i \\
-\sum_{i} m_i x_i y_i & \sum_{i} m_i (x_i^2 + z_i^2) & -\sum_{i} m_i y_i z_i \\
-\sum_{i} m_i x_i z_i & -\sum_{i} m_i y_i z_i & \sum_{i} m_i (x_i^2 + y_i^2)
\end{bmatrix},
\] where the eigenvector corresponding to the lowest eigenvalue indicates the major axis of the galaxy \( \rm M \). The orientation of the major axis of the galaxy in the XY plane can be defined as \[
\displaystyle \theta_{cen} = \arctan \frac{\rm M_{\rm y}}{\rm M_{\rm x}}.
\]
We visualise the results for one galaxy in Fig.~\ref{fig:central_orient_illustration_3snaps}, where the central galaxy $\log_{10}[\rm{M}_{\star} / \rm{M}_{\sun} ]= 11.96$ of a halo of mass  $\log_{10}[\rm{M}_{\rm {halo} } / \rm{M}_{\sun } ] = 14.02$ is shown in four snapshots with the indicated corresponding orientation of the major axis. All four panels show that the orientation of this galaxy does not change significantly in these timestamps. Figure~\ref{fig:central_orient_illustration2} shows the dynamical orientations at each snapshot for this galaxy. We note that the orientations at different times are mostly concentrated near the median orientation corresponding to 112 deg with a standard deviation of 35. In the very first times~(12-14 Gyr ago), the central galaxy had no specific orientation, but after the galaxy stabilised, its median orientation became 112 deg~(5-12 Gyr ago) with small fluctuations. Between 2 and 4 Gyr ago, this galaxy again did not have stable orientation of the major axis. The last 2 Gyr of the evolution of this galaxy were stable, and the median orientation during the last 3 Gyr: 87$\pm$6 deg and the vast majority of time stamps correspond to $z\sim$0 orientation. We estimate the number of abrupt changes in orientation~(the difference in orientation between two snapshots is greater than 30 deg). This galaxy has six abrupt orientation changes and four of them were in the first 2 Gyrs of galaxy life. Thus, the evolution of the orientation of each central galaxy could be described by the standard deviation from the median orientation and the number of abrupt changes.

We first discuss the number of abrupt(>30 deg between the next snapshots) for the central of the considered halos. We exclude the first 2 Gyr from this analysis since, in the first times of galaxy formation, they are not stable. Figure~\ref{fig:central_orient_illustration2} demonstrate the distribution of the number of abrupt changes in the orientation of the major axis: during all times of central galaxy life, they significantly change its dynamical orientation 5$\pm$1 times, and only 1$\pm$1 times in the last 3 Gyr.

\begin{figure}
    \centering
    \includegraphics[width=1\linewidth]{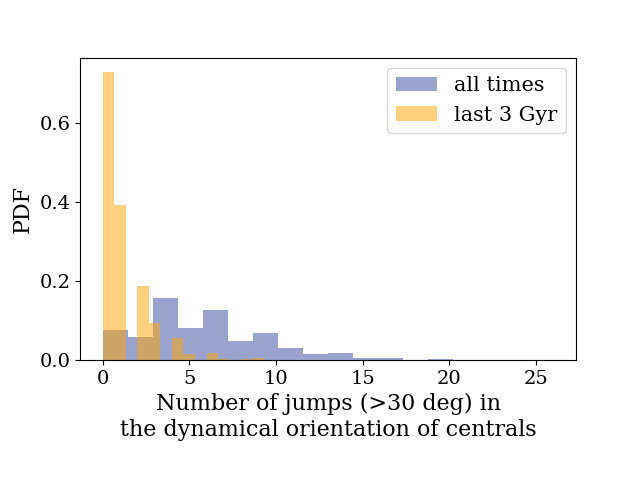}
    \caption{Number of abrupt changes~(the changes > 30 deg between two next snapshots) of the orientation of major dynamical axis for central galaxies in TNG100.}
    \label{fig:central_orient_jumps_number}
\end{figure}

\begin{figure}
    \centering
    \includegraphics[width=1\linewidth]{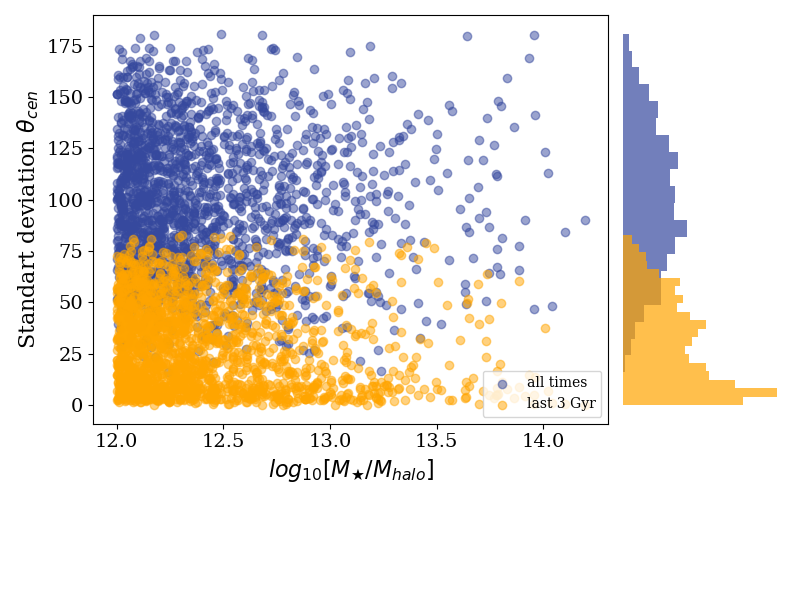}
    \caption{Standard deviation of dynamical orientation from median orientation for all times or last 3 Gyr ago as a function of halo mass.}
    \label{fig:central_orient_std_as_mhalo}
\end{figure}

Next, we check standard deviations from the median values of the orientation of the major axis for each galaxy. Figure~\ref{fig:central_orient_std_as_mhalo} demonstrated the standard deviation of orientation for each central galaxy during its life and in the last 3 Gyr as a function of the mass of the halo. We do not obtain any dependence on the halo mass. On average, during the entire life, the spread of the central galaxy of orientations is around 97$\pm$2, which corresponds to significant changes in orientation~(covering almost all possible orientations). However, this spread is significantly smaller if we consider only 3 last Gyr: 38$\pm$1\% of central galaxies have a standard deviation less than 15 deg and 38$\pm$1\% less than 30 deg and 58$\pm$2\% less than 30 degrees.

Combining the number of abrupt changes and standard deviations from the median values of the orientation of the major axis of the galaxy, we conclude that central galaxies significantly and often change their orientation during evolution. However, considering only the last 3 Gyr, we find that centrals rarely show abrupt changes in orientation and show a small standard deviation. This provides us with the possibility to assume that the orientation of the photometrical major axis used in the main part of the work is stable in the last 3 Gyr but not stable during the lifetime.

\section{The impact of the SFR definition on the presented results}
\label{app:sfr_def}

We examine here the sensitivity of the results presented in Figs.~\ref{fig:asgq_z0} and ~\ref{fig:asgq_yong_modulation_before_entrance} to the choice of star formation rate definition. While the main analysis employed instantaneous SFRs, we have now repeated it using 100 Myr averaged SFRs, as provided in the catalogues by \cite{Donnari+2019} and ~\cite{Pillepich+2019}, for all gravitationally bound galaxies. Aside from recalculating the main sequence between SFR and stellar mass, all other aspects of the analysis~(and code) remain unchanged.

\begin{figure*}
    \centering
    \includegraphics[width=1\linewidth]{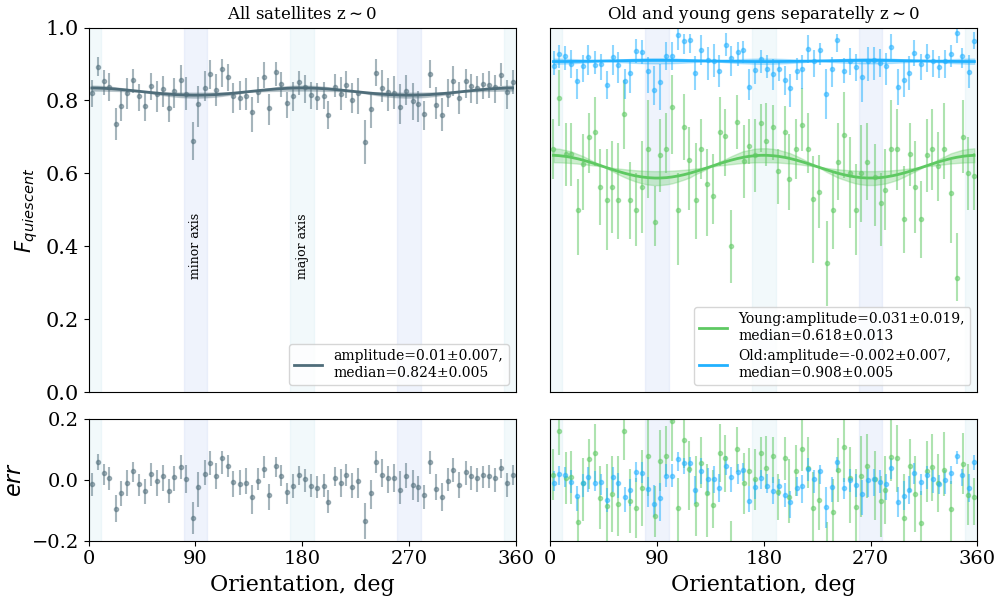}
    \caption{Same as Fig.~\ref{fig:asgq_z0} but with $\rm F_{\rm q}$ estimated using 100 Myr-averaged star formation rate definition.}
    \label{fig:asgq_z0_100myr}
\end{figure*}

\begin{figure*}
    \centering
    \includegraphics[width=1\linewidth]{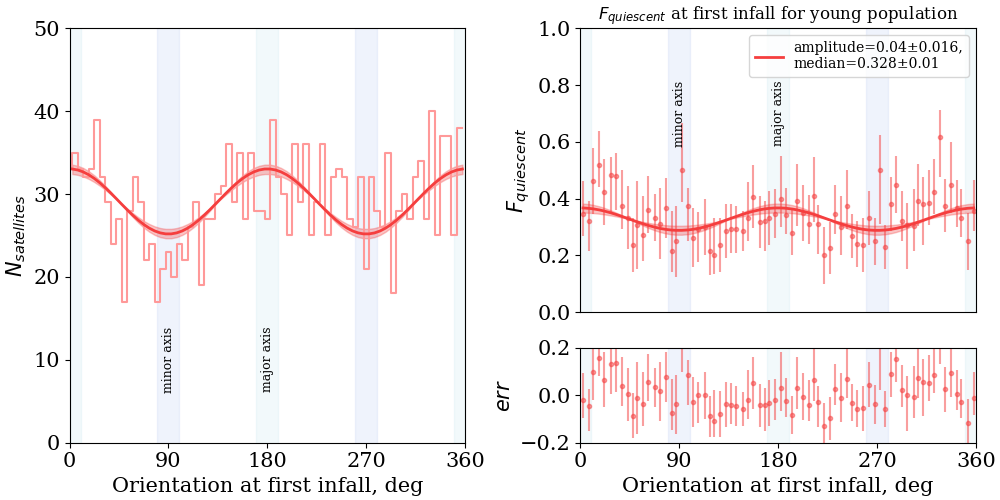}
    \caption{Same as Fig.~\ref{fig:asgq_yong_modulation_before_entrance} but with $\rm F_{\rm q}$ estimated using 100 Myr-averaged star formation rate definition.}
    \label{fig:asgq_yong_modulation_before_entrance_100myr}
\end{figure*}

The analogues of Figs.~\ref{fig:asgq_z0} and~\ref{fig:asgq_yong_modulation_before_entrance} concerning  100 Myr-averaged star formation rates are shown in Figs.~\ref{fig:asgq_z0_100myr} and~\ref{fig:asgq_yong_modulation_before_entrance_100myr}, respectively. We note a general trend: all median values of $\rm F_{\rm q}$ have increased. Next, Fig.~\ref{fig:asgq_z0} indicates that the modulation for all satellites at $z\sim0$ and for the young populations separately is less pronounced for the time-averaged star formation rates than for the instantaneous star formation rates, but still within 1$\sigma$ confidence intervals. The main conclusion remains the same: only young satellites are characterised by the ASGQ signal. Figure~\ref{fig:asgq_yong_modulation_before_entrance_100myr}  shows the presence of the modulation,  $\rm F_{\rm q}$, of the young population even before the first infall, also in agreement with Fig.~\ref{fig:asgq_yong_modulation_before_entrance}. Therefore, we conclude that the choice of the star formation definition does not affect our results.  

\section{ASGQ for groups and clusters separately}
\label{app:clusters_and_grousp_separately}

The quenching of satellite galaxies depends strongly on the host halo mass, which could introduce a bias in our wide mass range of host halos $12 < \log_{10} [\rm{M}_{\rm halo} / \rm{M}_{\odot}] < 14.2 $. Therefore here, we examine our main results from Figs.~\ref{fig:asgq_z0} and ~\ref{fig:asgq_yong_modulation_before_entrance} when groups $12 < \log_{10} [\rm{M}_{\rm halo} / \rm{M}_{\odot}] < 13.5 $ and clusters considered separately $13.5 < \log_{10} [\rm{M}_{\rm halo} / \rm{M}_{\odot}] < 14.2$.

Figures~\ref{fig:clusters_asgq} and \ref{fig:lowmass_asgq} below show the analogues of Figs. 3 and 4 from the original manuscript, but considering two different halo mass bins:  $12 < \log_{10} [\rm{M}_{\rm halo} / \rm{M}_{\odot}] < 13.5 $ and $13.5 < \log_{10} [\rm{M}_{\rm halo} / \rm{M}_{\odot}] < 14.2$. Overall, obtained  results are  more outstanding for cluster galaxies:
    \begin{itemize}
        \item The top panels of Figs.~\ref{fig:clusters_asgq}~(clusters) and \ref{fig:lowmass_asgq}~(groups) show that ASGQ exists only for massive clusters when all satellites are taken into account, regardless of their infall age~(comparing only left panels).  Nonetheless, when considering only the young generations of satellites (right panels), a modulation is recovered in both halo mass bins, even though it is still more pronounced in the case of clusters.   
        \item The bottom panels of Figs.~\ref{fig:clusters_asgq}~(clusters) and \ref{fig:lowmass_asgq}~(groups) show that clusters members have a more pronounced preferential orientation along the major axis of their central than group galaxies where the latter have almost no modulation at all. Moreover, clusters of galaxies have ASGQ even before entering their systems, unlike the group members. Besides, the median $F_{q}$ before infall significantly differs for clusters and the group's young satellites, revealing complex processes happen with galaxies before they enter their cluster. 
    \end{itemize}
Separating galaxies into low-mass groups and clusters would allow for a more detailed analysis. However, one of our primary goals was to ensure comparability with the results presented by \cite{Martin-Navarro+2021} to discuss their main conclusion, and such a division could complicate this comparison.

    \begin{figure*}
    \centering
    \includegraphics[width=1\linewidth]{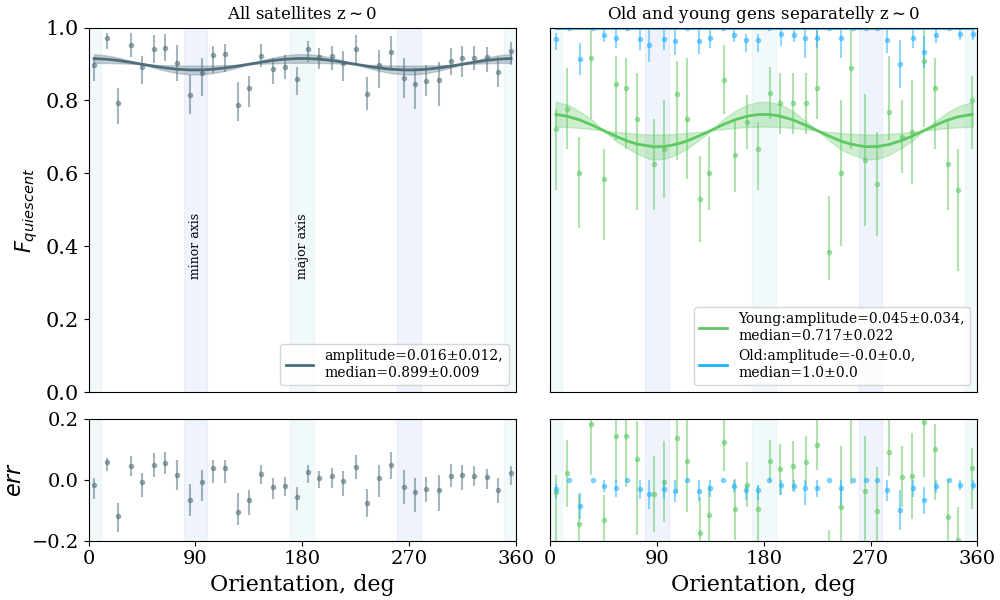}
    \includegraphics[width=1\linewidth]{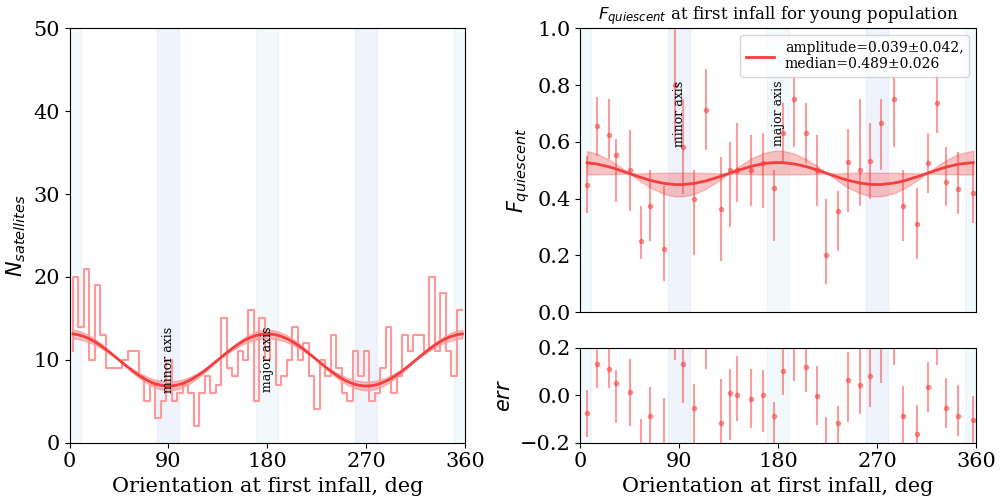}
    \caption{Analogues of Figs.~\ref{fig:asgq_z0}~(top panel) and~\ref{fig:asgq_yong_modulation_before_entrance}~(bottom panel) considering only galaxies in systems having a halo mass in the range $13.5 < \log_{10} [\rm{M}_{\rm halo} / \rm{M}_{\odot}] < 14.2$, typically corresponding to the cluster s regime. The size of bins increased in comparison with the original Figs due to a low number of statistics.  The dependencies were calculated for 46 clusters with 2721 satellites in total.  }
    \label{fig:clusters_asgq}
    \end{figure*}

    \begin{figure*}
    \centering
    \includegraphics[width=1\linewidth]{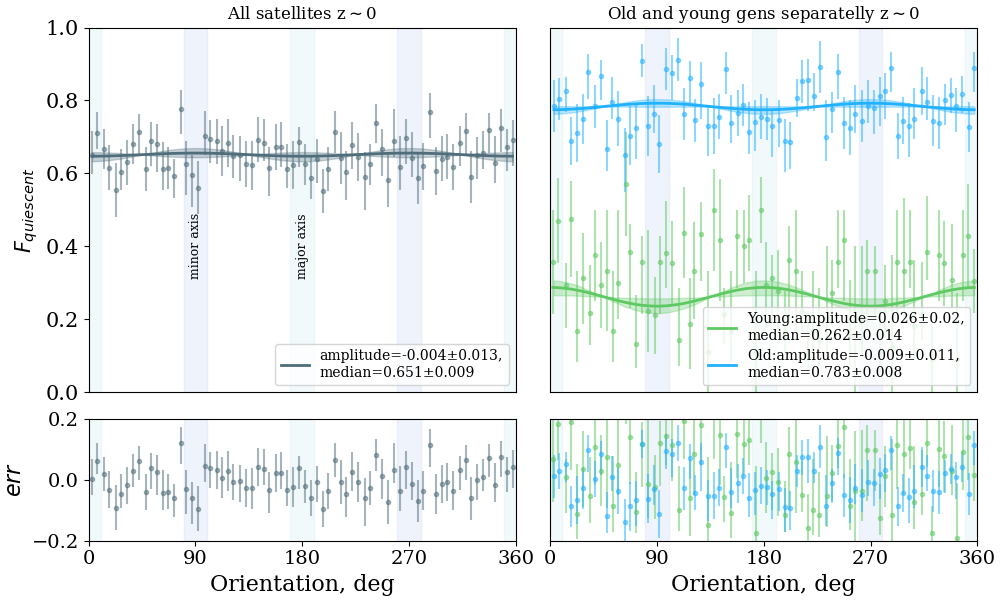}
    \includegraphics[width=1\linewidth]{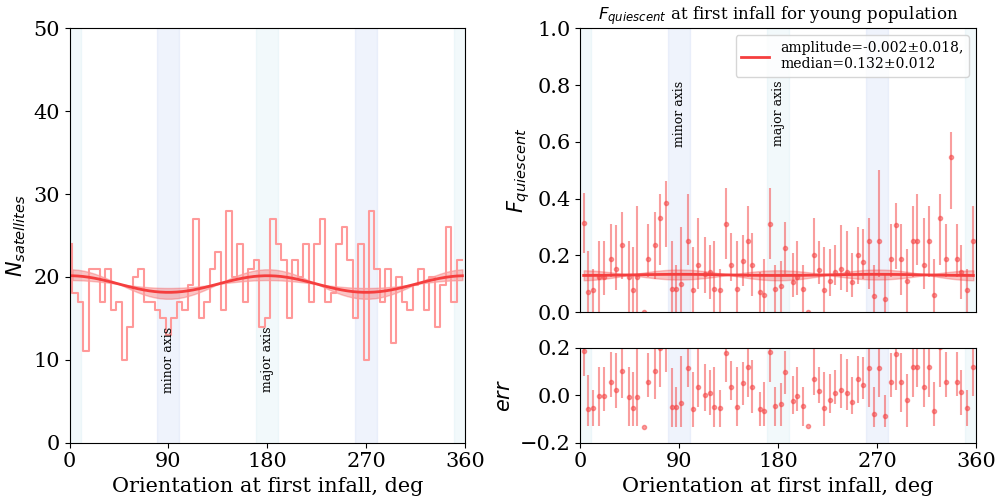}
    \caption{Analogues of Figs.~\ref{fig:asgq_z0}~(top panel) and~\ref{fig:asgq_yong_modulation_before_entrance}~(bottom panel) considering only galaxies in systems having a halo mass in the range  $12 < \log_{10} [\rm{M}_{\rm halo} / \rm{M}_{\odot}] < 13.5$, typically corresponding to the group regime. The dependencies were calculated for 1360 groups with 5446 satellites in total. }
    \label{fig:lowmass_asgq}
    \end{figure*}

\section{Parameter space analysis of fitting results}
\label{app:corner_plots}

In this appendix, we provide corner plots that illustrate the parameter distributions and correlations obtained from the MCMC fitting of Eq.~\ref{eq:fitting} for Fig.~\ref{fig:asgq_z0} and the left panel of Fig.~\ref{fig:asgq_yong_modulation_before_entrance}. Each plot contains details of the parameter estimation process with the following key features:
\begin{itemize}
    \item Diagonal panels show the one-dimensional marginalised distributions for each parameter a~(amplitude), b~(median $\rm F_{\rm q}$) and $f$~(rescaling term for errors) from Eq.~\ref{eq:fitting}, showing the probability density function and obtained values by vertical lines.
    \item Off-diagonal panels show the correlations between parameter pairs, with contours indicating credible regions at confidence levels of 1$\sigma$ and 2$\sigma$.
    \item The obtained values are indicated by black lines and displayed at the top of each column with 1$\sigma$ confidence intervals.
\end{itemize}

\begin{figure}
    \centering
    \includegraphics[width=1\linewidth]{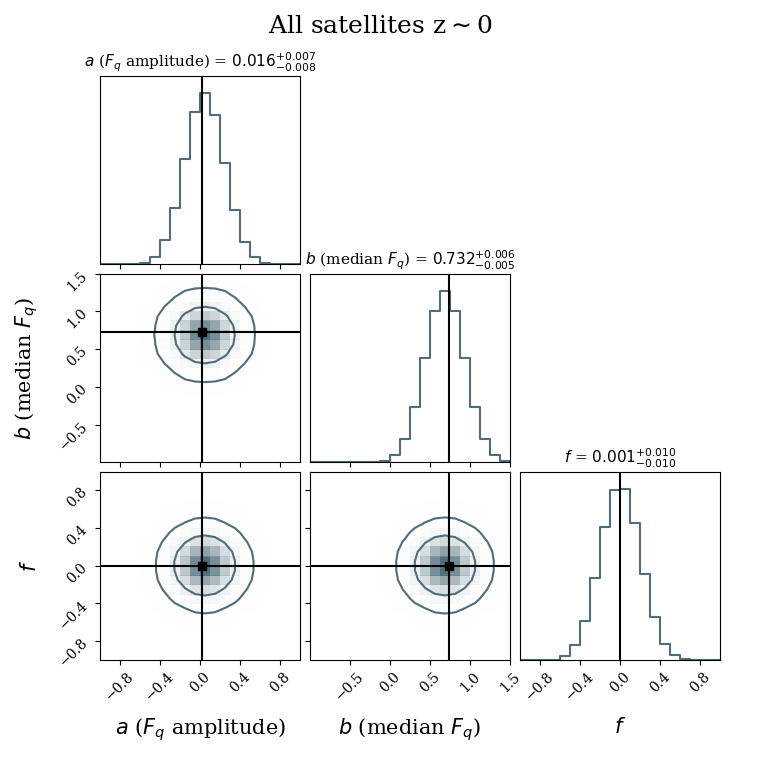}
    \caption{Parameter space analysis of fitting results of Eq.~\ref{eq:fitting} for all satellites at $z\sim0$ presented in Fig.~\ref{fig:asgq_z0}~(left panel). }
    \label{fig2_fitting1}
\end{figure}

\begin{figure}
    \centering
    \includegraphics[width=1\linewidth]{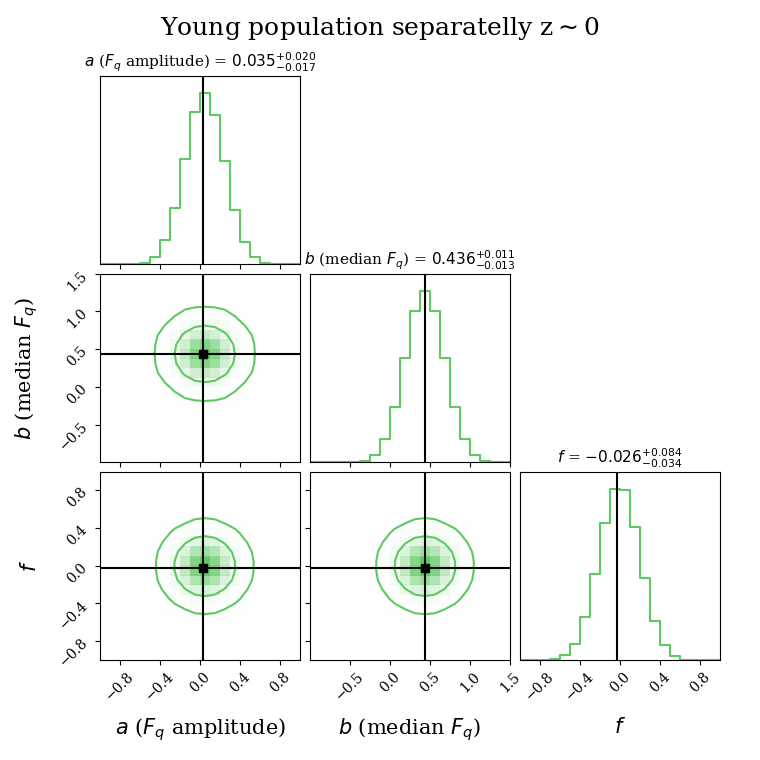}
    \caption{Same as Fig.~\ref{fig2_fitting1} but only for the young population satellites at $z\sim0$ presented in Fig.~\ref{fig:asgq_z0}~(right panel). }
    \label{fig:fig2_fitting_young}
\end{figure}

\begin{figure}
    \centering
    \includegraphics[width=1\linewidth]{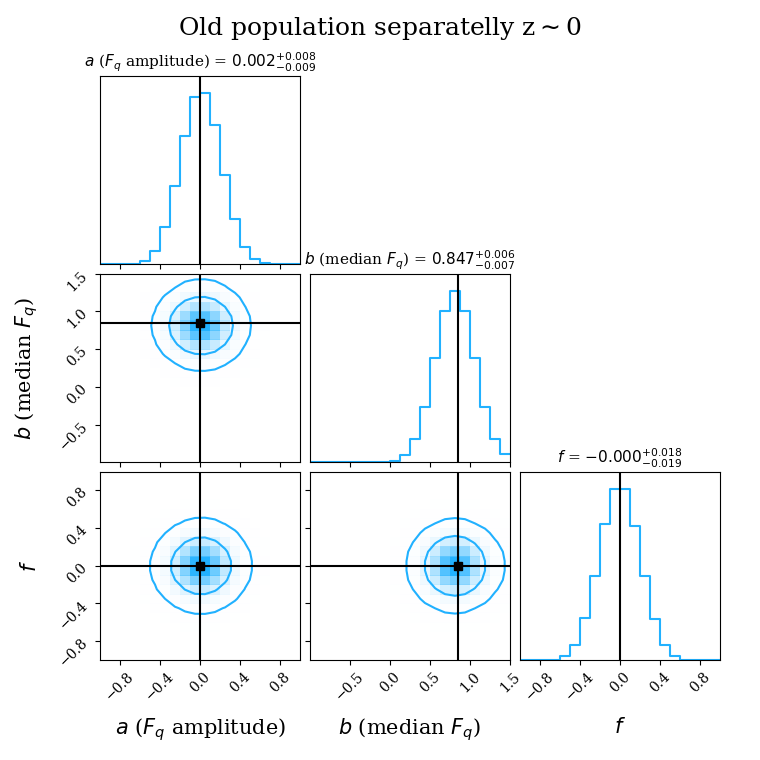}
    \caption{Same as Fig.~\ref{fig2_fitting1} but only for the old population satellites at $z\sim0$ presented in Fig.~\ref{fig:asgq_z0}~(right panel). }
    \label{fig:fig2_fitting_old}
\end{figure}

\begin{figure}
    \centering
    \includegraphics[width=1\linewidth]{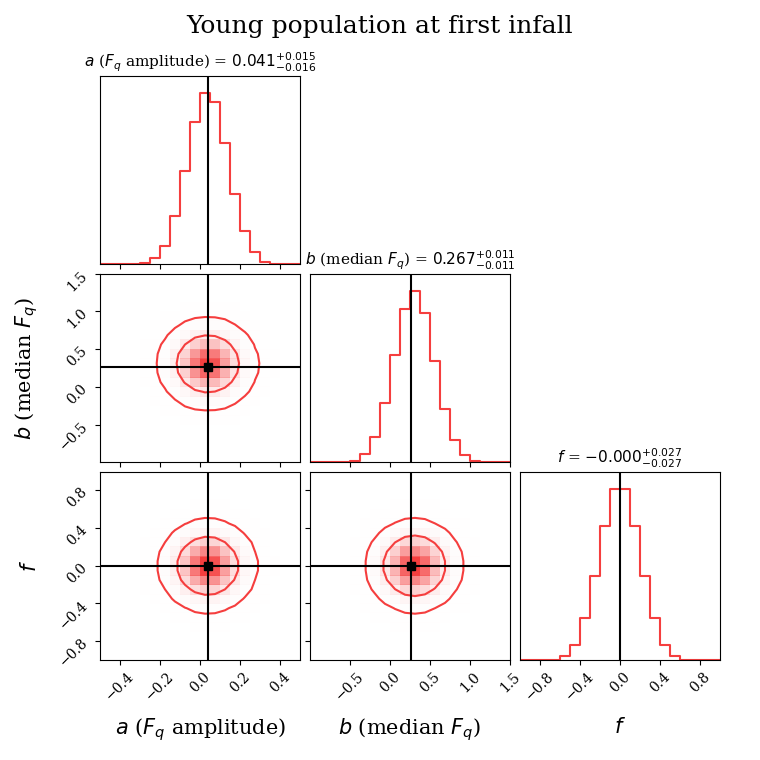}
    \caption{Same as Fig.~\ref{fig2_fitting1} but only for the young population satellites at first infall time corresponding to the right panel of Fig.~\ref{fig:asgq_yong_modulation_before_entrance}. }
    \label{fig:fig3_fitting}
\end{figure}

\end{appendix}

\end{document}